\documentclass[12pt]{elsarticle}
\usepackage{amssymb}
\usepackage{amsthm}
\usepackage{url}
\usepackage{boxedminipage}
\usepackage{color}
\usepackage{xcolor} 
\usepackage{alltt}
\usepackage{multicol}
\usepackage{hyperref}

\usepackage{ifthen}
\usepackage{bsymb,b2latex}
\usepackage{listings}
\usepackage{tikz}
\usetikzlibrary{matrix}

\usetikzlibrary{shapes,arrows}
\usetikzlibrary{patterns}

\tikzstyle{ground1}=[preaction={fill,top color=black!10,bottom color=black!5,shading angle=20}, fill,pattern=north west lines,minimum width=0.3,minimum height=0.6]
\tikzstyle{ground2}=[preaction={fill,top color=black!10,bottom color=black!5,shading angle=20}, fill,pattern=north east lines,minimum width=0.3,minimum height=0.6] 

\usepackage{amsmath,bm}

\usepackage{caption}

\theoremstyle{definition}

\def\iota{$\iota$}

\usepackage{lineno}

\begin{document}
	
	\begin{frontmatter}
		
		
		
\title{Mechanizing Operads with Event-B}
		

\author[]{Christian Attiogb\'e} 
\affiliation{organization={Nantes Université, École Centrale Nantes, CNRS}, 
			             addressline={LS2N, UMR 6004},
			            city={Nantes},
			             postcode={F-44000},
			             country={France}\\~\\  December 2025}
		

		
		\begin{abstract}
			
			Rigorous modelling of natural and industrial systems still conveys various challenges related to abstractions, methods to proceed with and easy-to-use tools to build, compose and reason on models.
			Operads are mathematical structures that provide such abstractions to compose various objects and garanteeing well-formedness.
			 
			Concrete implementations of operads will offer practical means to exploit operads and to use them for various technical applications. 
			Going from the mathematical structrures, we develop with Event-B a complete refinement chain that implements algebraic operads and their basic operations. 
	
			The result of this work, can be used from the methodological point of view to handle similar implementations for symbolic computation questions, and also to reason on symbolic computation applications supported by operads structures.
			
		\end{abstract}
		
		
		
		\begin{keyword}
			Operads \sep Semantics  \sep Symbolic computation \sep Formal modelling \sep  Event-B
			
		\end{keyword}
		
	\end{frontmatter}

	

	\section{Introduction}

Rigorous modelling of natural and industrial systems still conveys various challenges related to abstractions, methods to proceed with and easy-to-use tools to build, compose and reason on models.
Operads are mathematical structures that provide such abstractions to compose various objects and garanteeing well-formedness \cite{AlgebraicOperadsLoday2012}.

One interest in computing, is for structuring and modelling complex (component-based) systems and  their evolution. 
Consider for example the modelling of a manufacturing system and its further extension by replacing a module (operated by a human) by a machine-based module.

\paragraph{Motivation}
We are motivated by exploiting operads as practical systems modelling tools at the disposal of (software-based) system  engineering.
Indeed, formal modelling activities still need methods and generic libraries at the disposal of practitionners.
This challenge is advocated, for instance, in  \cite{doi:10.1098/rspa.2021.0099}.

\paragraph{Objectives}
The aim of this work is to implement libraries of generic structures and methods, in Event-B, to support formal  modelling and analysis.

\paragraph{Contribution}
The contributions of this article are manyfold: 
\textit{i)} a library of components to actually manipulate operads;
\textit{ii)} a rigorous framework for establishing manipulations of other similar algebraic structures.

\paragraph{Organisation}
The article is organised as follows. 
In Section \ref{section:backgroud-operads} we introduce the background on operads and on the Event-B method.
Section \ref{section:modelling} is dedicated to the modelling methodology.
In Section \ref{section:howto} we present how this work can be (re)used   
and
finally Section \ref{section:conclusion} draws some conclusions.

	\section{Background }
	\label{section:background}
This section provides an introduction to algebraic operads and to Event-B, the formalism that we use to model and mechanize the operads.
 
\subsection{Operads}
\label{section:backgroud-operads}

Operads are abstractions to encode algebraic structures and to consider the compositions of the relations on these structures.
Thus, operads allow to manipulate and generalize algebraic structures.
In the same way as functions  are applied to sets, morphisms are applied to operads; they are categories of objects.
There are many operads: topological operads, algebraic operads, modular operads, etc \cite{AlgebraicOperadsLoday2012}. In this work we focus on algebraic operads.

\paragraph{Algebraic operad}
A fundamental example is the \textit{endomorphism operad}, say $O_e = (\mathit{End}_X, \circ_i)$, defined on the set of functions  $\mathit{End}_X$, with the operations $\circ_i$,  where $X$ is a set and $i$ is a natural.

The operad $O_e = (\mathit{End}_X, \circ_i)$ is such that:

$\mathit{End}_X(k)~=~ X^k \to X$ with $k \ge 1$, is the set of the functions from $k$ products of $X$ to $X$

$\circ_i :  \mathit{End}_X(n) \times \mathit{End}_X(m) \to \mathit{End}_X(n+m-1)$  are the operations of composition of the functions on $\mathit{End}_X$, with two naturals $n$ and $m$, where $1 \le i \le n$.\\

Given $f\in \mathit{End}_X(n)$ and $g \in \mathit{End}_X(m)$, the operations $\circ_i$ are defined by
\begin{tabbing}
	\hspace{1.0cm}\=\hspace{0.5cm}\=\kill	
\> $\circ_i(f(x_1, x_2, \cdots, x_n), g(x_1, x_2, \cdots, x_m))) $\\
\>\> = $(f \circ_i g)(x_1, x_2, \cdots,  x_{n+m-1} )$\\
\>\> = $f(
x_1, 
x_2, 
\cdots,
x_{i-1}, 
g(x_i, x_{i+1}, \cdots, x_{i+m-1}),
x_{i+m},
\cdots, 
x_{n+m-1} )$ \\
\end{tabbing}

The function $id_X : X \to X$ is a neutral element of $o_i$. For  $f \in  X^n \to X$, with $n \ge 1$,  $f \circ_i id_X = f$ and $id_X \circ_1  f = f$. 

The functions of $\mathit{End}_X(0)$ (or  $X^0 \to X$)  correspond to the elements of $X$; they are thus constants (say $c \in X$) that can be used, given $f : X^n \to X$,  to block the composition at the given position $i$ in $f$; that is $f \circ_i c = c$, with $i\le n$.

Moreover the operations $\circ_i$ satisfy the following axioms:\\
\begin{tabular}{l}
	sequential composition: $(f \circ_i g) \circ_{i-1+j} h = f \circ_i (g \circ_j h)$ ~~~~if $1 \le i \le n$  and  $1 \le j \le m$ ~~~ \cr
	parallel composition: $(f \circ_i g) \circ_{k-1+m}  h =  (f \circ_k h) \circ_i g$ ~~~~if $1 \le i < k \le n$ \cr \cr
\end{tabular}

The tree operads (see Fig.~\ref{fig:treeoperad}) and the little cubes (see Fig.~\ref{fig:cubeoperad}) operads are famous graphical examples of representations (see \cite{sinha2010homologylittledisksoperad,AlgebraicOperadsLoday2012}) that provide an intuition on operad manipulations. In these examples, note that the elements of $X =\{x_1, x_2, \cdots \}$ are simply written 1, 2, 3, etc and the images or outputs of $f$, $g$, are simply denoted by a dot. This is an abstraction on $X$;  (formally the default unique output of an operad is abstracted with 0, and not written).

%

\usetikzlibrary{positioning}

\noindent
\begin{figure}[!ht]
		\centering
\begin{tikzpicture}[scale=0.7]
	\begin{scope}[xshift=0cm]
	\node {.} [grow=up]
	child {node {f}
		child {node {4}}
		child {node {3}}
		child {node {2}}
		child {node {1}}
	};
\end{scope}

\begin{scope}[xshift=3cm, yshift=1.5cm]
	\node {{\color{red}$\circ_2$}} [];
\end{scope}

\begin{scope}[xshift=5cm]
	\node {.} [grow=up]
	child {node {g}
		child {node {2}}
		child {node {1}}
	};
\end{scope}

\begin{scope}[xshift=7cm, yshift=1.5cm]
\node {=} [];
\end{scope}

\begin{scope}[xshift=10cm]
	\node {.} [grow=up]
	child {node {f}
		child {node {5}}
		child {node {4}}
		child {node {{\color{red}g}} [red]
			child {node {3}}
			child {node {2}}
		}
		child {node {1}}
	};
\end{scope}
\end{tikzpicture}
\caption{Composition of (tree) operads }
\label{fig:treeoperad}	
\end{figure}
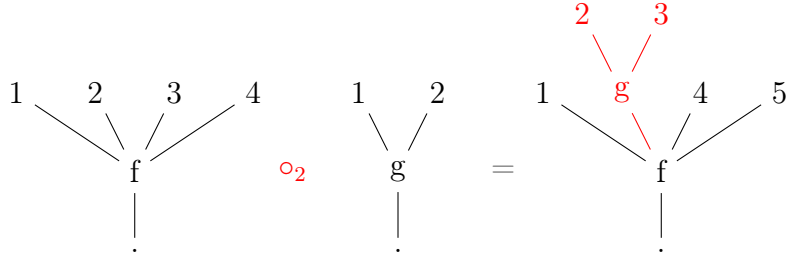

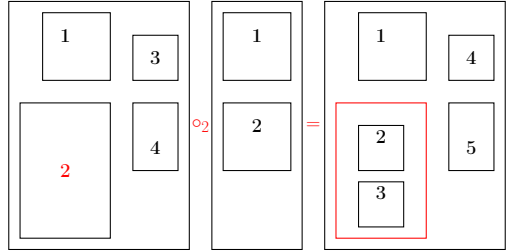
\begin{figure}
	\centering
	\begin{tikzpicture}[scale=0.3,transform shape]
	\draw (0.5,0.5) rectangle (8.5,11.5);
	\draw (1,1) rectangle (5,7) ;
	\draw (2,8) rectangle (5,11) ;	
	\draw (6,8) rectangle (8,10) ;
	\draw (6,4) rectangle (8,7) ;
	\node (1) at (3,10) {{\Huge \bf 1}} ;
	\node (2) at (3,4) {\color{red} {\Huge \bf 2}} ;
	\node (3) at (7,9) {{\Huge \bf 3}};
	\node (4) at (7,5) {{\Huge \bf 4}};

\begin{scope}[xshift=9.0cm, yshift=6.0cm]
	\node {{\Huge \bf {\color{red}$\circ_2$}}} [];
\end{scope}

\begin{scope}[xshift=9cm]
		\draw (0.5,0.5) rectangle (4.5,11.5);
	\draw (1,4) rectangle (4,7) ; 
	\draw (1,8) rectangle (4,11) ;	
	\node (1) at (2.5,10) {{\Huge \bf 1}} ;
	\node (2) at (2.5,6) {{\Huge \bf 2}} ;

\end{scope}	

\begin{scope}[xshift=14.0cm, yshift=6.0cm]
	\node {{\Huge \bf{\color{red}$=$}}} [];
\end{scope}

\begin{scope}[xshift=14cm]
	\draw (0.5,0.5) rectangle (8.5,11.5); 
	\draw [red] (1,1) rectangle (5,7) ; 
	\draw (2,8) rectangle (5,11) ;	
	\draw (6,8) rectangle (8,10) ; 
	\draw  (6,4) rectangle (8,7) ;
	\node (1) at (3,10) {{\Huge \bf 1}} ;
;
	\draw (2,4) rectangle (4,6); 
	\node (2) at (3,5.5) {{\Huge \bf 2}} ;
	\draw (2,1.5) rectangle (4,3.5);  
	\node (2) at (3,3) {{\Huge \bf 3}} ;
	\node (3) at (7,9) {{\Huge \bf 4}};
	\node (4) at (7,5) {{\Huge \bf 5}};
\end{scope}		
		
\end{tikzpicture}

\caption{Composition of (cube) operads}
\label{fig:cubeoperad}	
\end{figure}

\begin{figure}
	\centering
	\begin{tikzpicture}[scale=0.35,transform shape]
		\draw (0.5,4.5) rectangle (15.5,10.5);
		\draw (1,6) rectangle (4,9) ; 
		\draw (8,8.5) rectangle (11,10) ; 
		\draw [ground1] (4,7.5) rectangle (13,8) ;  
		\draw [ground2] (4,7.0) rectangle (13,7.5) ; 
	
		\draw (4.5,5) rectangle (8.0,6.5) ; 
		\draw (12.5,9.5) rectangle (15,8) ; 
		\draw (12.5,7) rectangle (15,5.5) ; 
								
		\node (1) at (2.5,8) {{\Huge \bf 1}} ;
		\node (1.1) at (2.5,7) {{\Huge  source}} ;
		\node (1.2) at (6.3,8.2) {\huge conveyor} ;
		\node (2) at (6,6) {{\Huge \bf 2}};
		\node (2.1) at (6.3,5.4) {{\Huge \bf machine}};
		\node (3) at (9.5,9.6) {\color{red} {\Huge \bf 3}} ;
		\node (33) at (9.5,9) {\color{red} {\huge human}} ;
		\node (4) at (13.5,9) {{\Huge \bf 4}};
		\node (41) at (13.5,10) {{\huge \bf products}};
		\node (4) at (13.5,6) {{\Huge \bf 5}};
		\node (4.1) at (13.5,5) {{\huge \bf rejects}};
		
		\begin{scope}[xshift=16.0cm, yshift=6.0cm]
			\node {{\Huge \bf {\color{red}$\circ_3$}}} [];
		\end{scope}
		
		\begin{scope}[xshift=16cm]
			\draw (0.5,9.0)rectangle (9.5,5.5);
			\draw (1,8.5) rectangle (3.2,6.0) ; 
			\draw (3.5,8.5) rectangle (6.0,6.0) ;	
			\draw (8.5,8.5) rectangle (6.5,6.0) ;	
			\node (1) at (2,7.5) {{\Huge \bf 1}} ;
			\node (1.1) at (2.0,6.7) {{\huge robot}} ;
			\node (2) at (4.5,7.5) {{\Huge \bf 2}} ;
			\node (3) at (7.5,7.5) {{\Huge \bf 3}} ;
			\node (3.3) at (7.5,6.7) {{\huge robot}} ;
		\end{scope}	
		
		\begin{scope}[xshift=26cm, yshift=6.0cm]
			\node {{\Huge \bf{\color{red}$=$}}} [];
		\end{scope}
		
		\begin{scope}[xshift=26cm]
		\draw (0.5,4.5)   [line width=0.8]  rectangle (21,12.5);
		\draw (1,6) rectangle (4,9) ; 
		\draw  [red] (7,8.5) rectangle (16,12) ; 
		\draw (8,9) rectangle (10.0,11.5) ; 
		\draw (10.5,9) rectangle (13,11.5) ; 
		\draw (13.5,9) rectangle (15.5,11.5) ; 
		
		\draw [ground1] (4,7.5) rectangle (18.5,8) ;  
		\draw [ground2] (4,7.0) rectangle (18.5,7.5) ; 

		\draw (4.5,5) rectangle (8,6.5) ; 
		\draw (17.5,9.5) rectangle (20.5,8) ; 
		\draw (17.5,7) rectangle (20.5,5.5) ; 

		\node (1) at (2.5,7.5) {{\Huge \bf 1}} ;
		\node (2) at (6,6) {{\Huge \bf 2}};
		\node (2.1) at (6.3,5.4) {{\Huge \bf machine}};
		
		\node (3) at (8.8,10.8) {\color{red} {\Huge \bf 3}} ;
		\node (3.3) at (9,10) { {\huge robot}} ;
		\node (4) at (11.5,10.8) {\color{red} {\Huge \bf 4}} ;
		\node (5) at (14.4,10.8) {\color{red} {\Huge \bf 5}} ;
		\node (5.1) at (14.5,10.0) { {\huge robot}} ;
		
		\node (6) at (18.5,9) {{\Huge \bf 6}};
		\node (6.1) at (19,10) {{\huge \bf products}};	
		\node (7) at (18.5,6) {{\Huge \bf 7}};
		\node (7.1) at (19.0,5) {{\huge \bf rejects}};

		\end{scope}		
		
	\end{tikzpicture}
	
	\caption{Modelling a manufacturing system via a composition of operads}
	\label{fig:MESoperad}	
\end{figure}
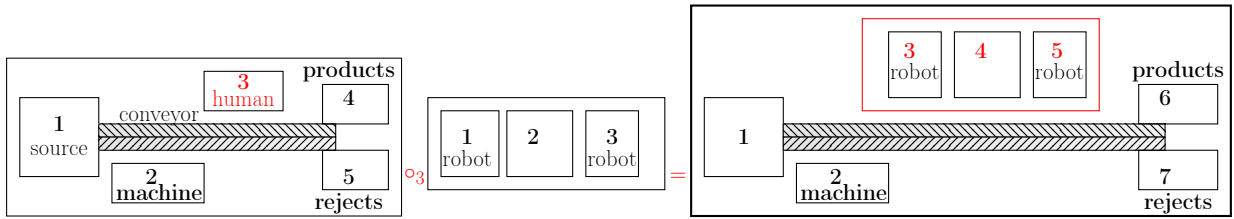

To illustrate the composition operator $\circ_i$, let us consider the following example.  
A preliminary architecture of a manufacturing system (see Fig.~ \ref{fig:MESoperad}) is modelled with an operad with 5 arguments that model 5 components of the manufacturing system, linked with a conveyor.
The argument 1 models a source of raw items; 
argument 2 models a module in which a machine applies a specific process  to the items picked from the conveyor. 
Argument 3 models a work station where a human picks an item from the conveyor,  applies a given process and deposits the item back to the conveyor. 
Finally the final products reach a store modelled as argument 4, and rejected products arrive at the store modelled as argument 5.  

Then, we would like to replace the module 3 operated by a human operator, in this architecture of the  manufacturing system,  by a new module made of 2 robots related to a machine; the first robot collects items from the conveyor and deposits them on a working table (modelled with component 2 inside the new module); after the items are processed, the second robot deposits them on the conveyor ; these three new components together, replace the human operator (see Fig. \ref{fig:MESoperad}). 

A composition of the operads, that of the initial system and an operad modelling the new three components, results in the new  architecture of the manufacturing system which can then be analysed before its effective implementation.     


\paragraph{Algebra of operads}

An algebra  over an algebraic operad $\mathit{P} =(End_X, \circ_i)$ is given by a set $X$, and a morphism of operads  $\mu : \mathit{P} \to {\mathit End}_X$. 
That means, for each operad $f_n : X^n \to X$ of $\mathit{P}$ with $n \ge 1$,  we have a morphism ($\mu(f) = \bar{f}$), such that $\bar{f_n} : X^n \to X$ encodes $f$.\\

We have only provided here a short background, more development and works on operads can be found in \cite{AlgebraicOperadsLoday2012},  \cite{whatisOperadStasheff2004}, \cite{idrissi:hal-03997161} for examples. \\

In the following we will build a generic model with Event-B, to encode operads together with the operations $\circ_i$.
Note that compared to the operads as depicted in Fig.~\ref{fig:treeoperad},  the graphical view of operads (see Fig.\ref{fig:cubeoperad}) is a view where details are added. As we will see in the introduction to Event-B and in the modelling with Event-B, adding details to an abstract model is done with \textit{refinement} technique.

\subsection{Event-B for system engineering}
Event-B \cite{EventB_Abrial2010,DBLP:journals/scp/HoangKBA09} is a modelling and development method where components are modelled as \textit{contexts} and \textit{abstract machines} which are composed and refined into concrete machines called \textit{refinements}.
In the Event-B modelling process, abstract machines constitute the dynamic part whereas \textit{contexts} are used to describe the static part.  

 A \textit{context} is made of carrier sets used as types. It may also contain constants ($c$), assumed properties (defined on the sets and constants) used as axioms ($Ax$),  and theorems ($Th$) derived from axioms.
A context can be extended to build larger contexts. 
Contexts are used (seen) by machines of a modelling project.  

An \textit{abstract machine}  describes a mathematical model of a \textit{system behaviour} seen as a discrete transition system.
Formally, an abstract machine is described by a state space made of typed variables ($x$) and invariants ($I(x,c)$), together with several \textit{event} descriptions. 
It can see one or several predefined contexts;
it can be extended to build larger machines.  The variable $x$ is used here for the list of state variables.

 \paragraph{State space of a machine}
 The variables constrained by the invariants (typing predicates, properties) describe the  state space of a machine.
 The transition from one state to the other is due to the effect of the events of the machine. Specific properties required by the model may be included in the invariant. The predicate $I(x,c)$ denotes the invariant of machine, with $x$ the list of state variables and $c$ the defined constants.

 \paragraph{Events of an abstract machine}
 Within Event-B, an event is the description of a system transition. 
 Events are  spontaneous and show the way a system evolves. 
 An event $e$ is modelled as a  \textit{guarded substitution}: 
 $e \triangleq (eG \Longrightarrow eB$) where $eG$ is the event \textit{guard} and $eB$ is the event \textit{body} or \textit{action}.
 It is described with the concrete form: 
 
 \centerline{$e$ $\triangleq$ \textrm{\textbf{any}} $lv$ \textrm{\textbf{where}} $eG(x, lv, c)$ \textrm{\textbf{then}} $eB(x,lv,c)$ \textrm{\textbf{end}}.} 
 
 $lv$ are local variables (or the parameters) of the event. To shorten the notation, $c$ is now forgotten in $eG(\cdots)$, $eB(x)$, $I(\cdots)$, $BA(\cdots)$.\\
  
 There is a specific initialisation event; it does not have a guard; it has the form:
 
 \centerline{ \textrm{init}  $\triangleq$  \textrm{\textbf{begin}} $eB(x)$ \textrm{\textbf{end}}}

 An event may occur only when it is enabled: its guard holds. A nondeterministic choice is made when several events are enabled. If no event is enabled,  the system is deadlocked.
 The action of an event $e$  describes, with simultaneous generalised substitutions  ($S$), how the system state evolves when this event occurs: disjoint state variables are then updated simultaneously.

The substitutions $S$ may be deterministic or nondeterministic.
For instance a basic deterministic substitution\footnote{Note the use of the ascii font to denote the Event-B code}  \texttt{x := E(x,c)}  is logically equivalent to the before-after predicate \textit{x' such that} ${x' = E(x,c)}$\footnote{$x'$ denotes the value of $x$ after the action}. This is symbolically written $ x' : (x' = E(x,c))$ where $x'$ corresponds to the state variable $x$ after the substitution and $E(x,c)$ is an expression. 
Nondeterministic substitutions are expressed as $x:\in E$ where $E$ is an expression or as $x~:|~P(x, c)$ where $P(x, c)$  is a predicate which characterises the possible values of $x$.

In Event-B \textit{proof obligations} are defined to establish the model consistency via the invariant establishement by the initialisation and the invariant preservation by the other events (that change the state variables).
Specific properties (included in the invariant) of a system are also proved in the same way.
Formally the schema of the proof obligations are the following sequents:

Initialisation proof obligation \hfill $BA_{\textrm{init}}( x') \vdash I(x')$

Preservation proof obligation \hfill $I(x), eG(x,lv), BA_e(x, x') \vdash  I(x')$

\paragraph{Refinement} An important feature of the Event-B method is the availability of refinement technique to design more and more concrete machines and systems from the abstract model(s), by stepwise enrichments of the abstract model. During the refinement process new variables ($y$) are introduced; the invariant is strengthened without breaking the abstract invariant $I(x,c)$, and finally the event guards are strengthened. In the invariant $J(x,y)$ of the refinement, abstract variables ($x$) and concrete variables ($y$) are linked. The refinements are also accompanied with proof obligations in order to prove their correctness with respect to the abstract model they refine.
The proof obligation of an event with a before-after predicate $BAA(x,x')$ in the abstract machine model and a before-after predicate $BAC(y,y')$ in the concrete (refined) model is: $$I(x) \land J(x,y) \land BAC(y,y') \implies \exists x'.(BAA(x,x') \land J(x',y'))$$ The intuition is that a concrete event can be simulated by an abstract one.

\paragraph{\texttt{Rodin} Tool} \texttt{Rodin}\footnote{{\small http://wiki.event-b.org/index.php/Main\_Page}} \cite{rodin_AbrialBHHMV10} is an open and  extensible tool, structured with several modules (plug-ins), dedicated to building and reasoning on B models. \texttt{Rodin} integrates various theorem-provers and also the \textsf{ProB} model-checker~\cite{LeuschelButler:FME03}.

	\section{Modelling Operads with Event-B}
	\label{section:modelling}
We have to define Event-B machines to model operads, structured as $O_X = (\mathit{End}_X(n), \circ_i)$, with a set $X$ as a parameter, $n$ a natural, $\mathit{End}_X(n)$ the set of functions of $X^n \rightarrow X$ and $\circ_i$ the composition functions parameterised by a natural $i$.
The parameter $X$ is a fixed set, it will be defined in a context;  
$\mathit{End}_X(n)$ depends on an parameter $n$ for each of its element which is an operad with $n$ input arguments, thus $\mathit{End}_X(n)$ will be constructed dynamically with an event;  
$i$ is a parameter of a composition, it will also be defined dynamically, together with the two operads, parameters of $o_i$.

The composition of operads involves the evolution of the Event-B model, it then will be modelled with an event, namely \textsf{composeSeq}.
For the sake of brievety, we use the notation $f(1,2,3,4)$, $g(1,2)$ and $f(1, g(2,3), 4, 5)$ for the operads in Fig.~\ref{fig:treeoperad}. This is a first abstraction in the modelling, as we do not use yet the set $X$ but natural numbers. The default output argument is also neglected (it is abstracted as 0 and appears simply as a dot). 
The input arguments $1,2,3,4,\cdots$ are refered to as \textit{positions} or \textit{labels} in their operad. Then, we compare and relabel them in the composition.

\subsection{Encoding operads structure}
The structure of operads as defined in  Section~\ref{section:background} is partially  encoded in an Event-B  context.\\
\begin{figure}
	\begin{center}
		\begin{boxedminipage}{15cm}
			\input{ctx0_2309.tex}
		\end{boxedminipage}
	\end{center}
	
	\caption{Context for the modelling}
	\label{figure:Context0}
\end{figure}

We have introduced a context (see Fig.~\ref{figure:Context0}) for the necessary modelling ingredients or parameters, and an abstract machine that contains four variables (\textit{myOperads, arityOp, foliage and outOp}) and an event (\texttt{newOperad}) to built new operads, and one event (\texttt{composeSeq}) to compose two given operads.\\
\begin{center}
	\begin{boxedminipage}{15cm}
		\SingleHeader{Operads\_Abs}
\MACHINE{Operads\_Abs}{}{}{Abstract of operads}
\VARIABLES{
	\Variable{$\cdots$}{}
}
\INVARIANTS{
		\Invariant{inv10}{false}{$myOperads \subseteq{} OPERAD$}{// all defined operads}
	\Invariant{inv30}{false}{$arityOp \in{} myOperads \pfun{} (1\upto{}maxFol)$}{ // the arity (n, m, ...)}
	\Invariant{inv40}{false}{$foliage \in{} ((1\upto{}maxFol) \rel{} myOperads)$}{// foliage of each operad}
	\Invariant{inv60}{false}{$outOp \in{} myOperads \pfun{} \pow{}(1\upto{}maxArgs)$}{ // output of operads }
}
$\cdots$
\END 
	\end{boxedminipage}
\end{center}

As introduced in Sect.~\ref{section:backgroud-operads} any operad $f$ has a given number ($n$)  of inputs and one output by default. The inputs are ordered, and their positions are used in the composition of operads. 
To generalise the modelling and its future extension, we consider the function $outOp$ that may define more than one output, but it is constrained it the current model.

In the simplest case, the composition of two operads $f(1..n)$ and $g(1..m)$, results in a new operad with $(n+m-1)$ ordered inputs and 1 output. Practically, we have a set of operads (denoted $myOperads$) which is growing after each new added operad. Roughly, the variable $myOperads$ models the operads $End_X$. 

In the general case, that is a  composition at any position after previous compositions,  modelling the composition operation is more complex.
Consider for example the three operads $f(1, 2, 3, 4)$, $g(1, 2, 3)$ and $h(1, 2, 3)$,
the composition $(f \circ_2 g)$ results in an operad with 6 leafs if we consider the tree representation (Fig.~\ref{fig:treeOperadFG}); the composition $((f \circ_2 g) \circ_4 h)$ results in an operad with 8 leafs (see Fig. \ref{fig:treeOperadFGH}). But, note that $h$ is inserted in $g$ according to the position (4) of the compostion, whereas the insertion will be done in $f$ if we have to compose at the position  5 for example: $((f \circ_2 g) \circ_5 h)$.

	\begin{minipage}[b]{8cm}
			\centering
			\begin{tikzpicture}[scale=0.6]
				\begin{scope}[xshift=10cm]
					\node  {.} [grow=up]
					child {node {f}
						child {node {6}}
						child {node {5}}
						child {node {{\color{red} g}} [red]		
							child {node {4}}
							child {node {3}}
							child {node {2}}
						}
						child {node {1}}
					};
				\end{scope}
			\end{tikzpicture}
			
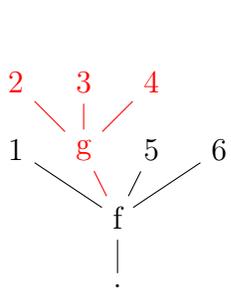
\captionof{figure}{Composition $(f \circ_2 g)$}
			\label{fig:treeOperadFG}	
	\end{minipage}
\begin{minipage}[b]{8cm}
		\centering
\begin{tikzpicture}[scale=0.6]
	\begin{scope}[xshift=10cm]
		\node  {.} [grow=up]
		child {node {f}
			child {node {8}}
			child {node {7}}
			child {node {{\color{red}g}} [red]
				child {node {{\color{blue}h}} [blue]
					child {node {6}}
					child {node {5}}
					child {node {4}}
				}
				child {node {3}}
				child {node {2}}
			}
			child {node {1}}
		};
	\end{scope}
\end{tikzpicture}

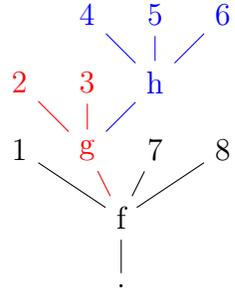
\captionof{figure}{Composition $((f \circ_2 g) \circ_4 h)$}
\label{fig:treeOperadFGH}	
\end{minipage}
	

Therefore, given two operads $op_1$ and $op_2$ and a position $i$ in $op_1$,  depending on the paramater $i$, the composition is achieved directly in $op_1$ (if the value of $i$ is up to the arity of $op_1$ and if a previous composition did not yet restructure $op_1$) or in an operad previously inserted in $op_1$. Moreover, the arities of the used operads, the relabelling of the operads's arguments (for instance in $h(4,5,6)$), the number of final leafs,  do not help directly to determine the operad in which the composition is achieved.

To master this complexity, we have introduced some abstractions in the modelling of operads and their compositions:  
\begin{itemize}
	\item an operad has a \textit{foliage}: it is all its input positions (as the leafs), whatever the level of composition; for instance $\{1, 2, 3, 4, 5, 6, 7, 8 \}$ is the foliage of the operad $r = ((f \circ_2 g) \circ_4 h)$.\\
	There is a total function from $\{1, 2, 3, 4, 5, 6, 7, 8 \}$ to $r$ (in green in  Fig.~\ref{fig:treeOperadFGH-foliage}). The cardinal of the foliage is greater or equal to the arity of the operad. 
	\item after a composition, an operad is hooked in another one; for instance in Fig. \ref{fig:treeOperadFGH}, the operad $h$ is hooked in the operad $g$, the operad $g$ is hooked in $f$.
	\item a given position $i$ has a \textit{hat operad} ; $i$ is directly in an operad; for instance the hat of the position $3$ is the operad $g$, the hat of the position $5$ is the operad $h$.
	\item after a composition into an elementary operad $op$, the relation between its positions or arguments and $op$,
	is no more a total function, as some positions are deleted (those use for the composition), and others relabelled. This prevents for using a total function to link the positions to their operads.
\end{itemize}

\begin{figure}
	\centering
	\begin{tikzpicture}[scale=0.6]
	
		\begin{scope}[xshift=10cm]
			\node (ra) {.} [grow=up]
			child {node {f}
				child {node (8) {8}}
				child {node (7) {7}}
				child {node {{\color{red}g}} [red]
					child {node {{\color{blue}h}} [blue]
						child  {node (6) {6}}
						child {node (5)  {5}}
						child {node (4) {4}}
					}
					child {node (3) {3}}
					child {node (2) {2}}
				}
				child {node (1) {1}}
			};
			\draw [dashed,->, green] 
			(8) to [bend left = 30] (ra) 
			(7) to [bend left = 40] (ra) 
			(6) to [bend left = 60] (ra) 
			(5) to [bend left = 40] (ra) ; 
			\draw [dashed,->, green ] 
			(4) to [bend right = 30] (ra)
			(3) to [bend right = 20] (ra)
			(2) to [bend right = 15] (ra)
			(1) to [bend right = 30] (ra)
			;
		\end{scope}
	\end{tikzpicture}
	\captionof{figure}{Composition $((f \circ_2 g) \circ_4 h)$}
	\label{fig:treeOperadFGH-foliage}	
\end{figure}

These observations justify the  structures we have used to model the operads (see Fig.~\ref{figure:M1_invariant}); note that to ease the reading and avoid data duplications, we only give an excerpt of the refinement. However, the full architecture of the development  is given as a synthesis in Section~\ref{section:archiDevEventB}. 

\begin{figure}[!h]
	\begin{center}
		\begin{boxedminipage}{15cm}
        \input{V6_invariantM1.tex} 
       \end{boxedminipage}
    \end{center}

\caption{Excerpt of operads structuring in Event-B}
\label{figure:M1_invariant}
\end{figure}

\subsection{Modelling the composition of operads}
First, we introduce an event that models the construction of an elementary (non-compound) operad. The occurrences of this event generate a set of elementary operads for subsequent compositions.
\paragraph{Construction of a new operad}
We build a new operad from two parameters: its numbers of inputs $rr$ and outputs $vv$. The parameter $rr$ is used to generate a sequence of $rr$ naturals ($1..rr$). The event \texttt{newOperad}  (see Fig.~\ref{figure:M1_newOperad}) gives the modelling of the introduction of a new operad, for instance $f(1,2, 3, 4)$. 
The new operad created by the event is added to \textit{myOperads} and may be used in future compositions.
Note that, $inOp, outOp$ and $myOperads$ are properly updated. For the sake of brievety, we merge here actions (labelled \texttt{$a_i$}) of the abstract machine  and those (labelled \texttt{$ra_i$}) of the refinement; the same holds for the guards  (labelled with \texttt{$g_i$} for the abstract machine, and with $rg_j$ for the refinement).

\begin{figure}[!h]
	\begin{center}\noindent
		\begin{boxedminipage}{16.3cm}
		\hspace{0.5cm}	
	\input{V6_M1_newOperad.tex} 
\end{boxedminipage}
\end{center}

\caption{Excerpt of the \textsf{newOperad} event}
\label{figure:M1_newOperad}
\end{figure}

\paragraph{Composition of operads}
Modelling the composition of operads is very challenging due to the multiple involved cases and the related restructuring of the parameter operads. 
After analysis and generalisation to any two operads $op_1$ and $op_2$, either elementary or resulting from previous compositions, we identify four different cases depending on the two operads to be composed and the position named here $ii$ of the composition.  In the required conditions of functionning, the variable $ii$ is considered in the foliage of $op_1$. The operad $op_2$ has its foliage, say \textit{foliage2}.
\begin{itemize}
\item the position $ii$ in the foliage is directly in the operad $op_1$ (not in an internal operad that composed $op_1$);
this position $ii$ will be deleted from the positions of $op_1$. 
\begin{itemize}
	 \item all the arguments at positions which are  little than $ii$ remain in the foliage of the composition result, without any changes.
	
	\item all the arguments at the positions greater than $ii$ remain in the foliage of the composition result, but they should be relabelled, by increasing their current label with the cardinal of \textit{foliage2} minus 1. The input positions in $op_1$ should be modified by considering their new labels.
\end{itemize}
\item the position $ii$ is indirectly in one of the internal operads  that composed $op_1$ (the operads already hooked in $op_1$).
\begin{itemize}
\item all the positions little than $ii$ remain  in the foliage of the composition result, and remain in their initial hat operad,
\item all the positions greater than $ii$ remain in the foliage of the composition result, but they should be relabelled, by increasing their current label with the cardinal of \textit{foliage2} minus 1.
\end{itemize}
\end{itemize}

In all the cases, the foliage of $op_2$ should be totally relabelled, by adding $ii\!-\!1$ to each position in \textit{foliage2}, whatever their \textit{hat operad} within $op_2$.
 
For the modelling, we then need to identify the \textit{hat operad}, given a composition parameter $ii$, and achieve the updates accordingly, in $op_1$ and in $op_2$. We also need to identify and modify the operads hooked in others, whatever their position.

\begin{figure}[!ht]
	\begin{center}
		\begin{boxedminipage}{16.3cm}
			\indent
			\input{V6_M1_composeOperads.tex} 
		\end{boxedminipage}
	\end{center}
	
	\caption{Excerpt of the event \textsf{composeSeq}: the action part}
	\label{figure:M1_composeOperads}
\end{figure}

The event \texttt{composeSeq} (see Fig.~\ref{figure:M1_composeOperads}) has several parameters among which: $op1$, $ii$, $op2$.
Its results is the modification of the operad $op1$, in which the output of $op_2$ replaces the input at position $ii$ in $op1$, and also with the full restructuring related to the composition.

Formally, the position $ii$ which should be in the foliage of $op_1$, is first used to identify the operad in which the composition will actually be done. The relation $GHatOp$ (see Fig.~\ref{figure:M1_invariant}, \texttt{invr20}) is used for this purpose. 
After each composition, relabelling the arguments of the operad, is done in the foliage for all arguments greater than $ii$.  

We structure the event \texttt{composeSeq}  with three main parts in its guards. In the first part we introduce the required variables, their typing constraints and some intermediary computations in local variables; for instance the operads already hooked in $op1$, to prepare further conditions in the guard (see Fig. \ref{figure:M1_guard1}). The second and the third guard parts consider nondeterministic valuations of intermediary variables, used to anticipate the simultaneous changes of state variables in the action part of the event. We provide (reshaped) excerpts of the guards (see Fig.~\ref{figure:M1_guard1}, Fig.~\ref{figure:M1_guard2} and Fig.~\ref{figure:M1_guard3}).

\begin{figure}[!h]
	\begin{center}
		\begin{boxedminipage}{16.3cm}
			\input{V6_M1_guard1.tex} 
		\end{boxedminipage}
	\end{center}
	
	\caption{Part1 of the guard of event composeSeq}
	\label{figure:M1_guard1}
\end{figure}

Note the intensive use of relational operators that help in modelling the transformation of the used set and relational strutures: the operators $\Union(\cdots)|\{\cdots\}$ and $A \bunion B \bunion \cdots$ are respectively the quantified and generalised union;
$s \domres r$ is the restriction of the domain of a relation $r$ to the set $s$,  
$r \ranres{} s $ is the restriction of the range of a relation $r$,
$s \domsub{} r $ is the substraction of $s$ from the domain of the relation $r$,
$r \ransub{} s $ is the substraction of $s$ from the range of $r$.
$r_1 \ovl{} r_2$ overwrites a relation $r_1$ by the left, with $r_2$.\\

The actions of the \texttt{composeSeq} event (see Fig.~\ref{figure:M1_composeOperads}) specify that:\\ 
(\textsf{a1}) $op2$ is deleted from the domain of $outOp$, since it is now used in the composition; \\
(\textsf{a2}) the foliage relation is updated to take account of the new foliage, linked to $op1$.\\
(\textsf{ra1}) $op2$ is now hooked in $hatopii$, the operad found as the hat of the position $ii$;\\
(\textsf{ra2}) the operads hooked in $op2$ are now transitively hooked in $op1$, to have a global relation;\\
(\textsf{ra3}) the hat operads are updated at all levels wrt to the new labels of their arguments;\\
(\textsf{ra4}) input positions in $op1$ and $op2$ are  updated depending on the four identified cases. \\

In the second  part of the guard (Fig.~\ref{figure:M1_guard2}): 
\texttt{rg51} expresses that $hatop1$ is all the hat operads in $op1$;
with the guard  \texttt{rg82} expresses that  $hatlii$ is the hats of the position $uu$ little than $ii$;
the formula labelled \textsf{rg102}, expresses the update of the positions $kk$ in (the foliage of) $op1$, greater than $ii$, by increasing them by ($kk+(cardfol2-1)$). All these guards prepare the  update of the hat in op1 and op2, through the guard \texttt{rg119}. 

In the third part of the guard (Fig.~\ref{figure:M1_guard3}), we prepapre the conditions and constraints for the update of the input arguments of ($inOp$) of the operads $op1$ and $op2$; with the guard \texttt{rg122}, \textit{luuupdinop1} expresses the inputs $kk$ little than $ii$ which remain unchanged; with the guard \texttt{rg134}, \textit{giiupdinop1} expresses the update of the inputs $kk$ greater than $ii$. Both are then merged. The update conditions of the inputs of $op2$ are expressed through the guard \texttt{rg162}.  
\begin{figure}[!ht]
	\begin{center}
		\begin{boxedminipage}{16.3cm}
			\input{V6_M1_guard2.tex} 
		\end{boxedminipage}
	\end{center}
	
	\caption{Part2 of the guard of event composeSeq}
	\label{figure:M1_guard2}
\end{figure}

\begin{figure}[!ht]
	\begin{center}
		\begin{boxedminipage}{16.3cm}
			\input{V6_M1_guard3.tex} 
		\end{boxedminipage}
	\end{center}
	
	\caption{Part3 of the guard of event composeSeq:  }
	\label{figure:M1_guard3}
\end{figure}

\subsection{Emerged specific consistency properties}
To enforce the consistency and the correction of the composition operator (modelled as an event), we have introduced some specific consistency properties as follows.


\begin{center}
	\begin{boxedminipage}{16.3cm}
			\vspace{-0.5cm}
		\Invariant{SP1}{false}{$\\((myOperads \neq{} \emptyset{}  \land{} (gHookOp \neq{} \emptyset{}) \land{} (GHatOp \neq{} \emptyset{}))  \limp{}\\~         		(\forall{}op\qdot{}(op \in{} myOperads \land{} op \in{} ran(GHatOp) \land{} op \in{} dom(inOp) \land{} op \notin{} dom(gHookOp) \land{} (op \in{} ran(gHookOp)) \land{} op \in{} ran(foliage) ) \limp{}\\~         			(  (dom(dom(GHatOp \ranres{} \{op\})) \bunion{} inOp(op)) = dom(foliage \ranres{} \{op\}) )))$}{\\ all the args in the foliage have a hat}
	\end{boxedminipage}
\end{center}
The property formulated in the invariant (\texttt{SP1}) specifies that all the arguments that appear in the foliage have a hat operad; they are in the range of the $GHatOp$ relation which relates the operads hooked in other given ones and their foliage. The arguments in the foliage are such that they are transitively under the hat of the operads.

\begin{center}
	\begin{boxedminipage}{16.3cm}
			\vspace{-0.5cm}
			\Invariant{SP2}{false}{$\\(((myOperads \neq{} \emptyset{}) \land{} ((foliage) \neq{} \emptyset{}) \land{} (gHookOp \neq{} \emptyset{})  )   \limp{}\\~         	(\forall{}op\qdot{} ((op \in{} myOperads \land{} op \in{} ran(foliage) \land{} op \in{} ran(gHookOp)  \land{} op \in{} dom(inOp) ) \limp{}\\~         			( dom(foliage \ranres{} \{op\}) = ((\Union{}oo\qdot{}(oo \in{} myOperads \land{} oo \in{} (dom(gHookOp \ranres{} \{op\})) \land{} (oo \in{} dom(inOp)) \land{} (inOp(oo) \subseteq{} (1\upto{}maxFol)))| (inOp(oo)) ) \bunion{} (inOp(op)))  )~         		   ) ) )$}{\\the input arguments in a foliage of an operad op, are from the operads hooked in op}
	\end{boxedminipage}
\end{center}
The invariant  \texttt{SP2} specifies that the input arguments in a foliage of an operad $op$, that is $dom(foliage \ranres{} \{op\})$, come only from the operads hooked in $op$.

\begin{center}
	\begin{boxedminipage}{16.3cm}
			\vspace{-0.5cm}
		
	\Invariant{SP3}{false}{$\\(( (myOperads \neq{} \emptyset{}) \land{} (inOp \neq{} \emptyset{}) \land{} (arityOp \neq{} \emptyset{}) \land{} (hookOp \neq{} \emptyset{})  )   \limp{}\\~         \forall{}op\qdot{}((op \in{} myOperads \land{} op \in{} dom(inOp)  \land{} op \in{} dom(arityOp)\land{} op \in{} ran(hookOp) \land{} finite(hookOp \ranres{} \{op\}))~~         \limp{} ~~card(inOp(op)) = (arityOp(op) - card(hookOp \ranres{} \{op\}))))$}{\\if the inargs are empty, that is operads are hooked in them all}
	\end{boxedminipage}
\end{center}

Finally, invariant  \texttt{SP3} stipulates that if an operad no longer has any input arguments, this is the result of their replacement by compositions, by other operads (thus hooked at their position)

\subsection{Architecture of the Event-B development}
\label{section:archiDevEventB}

\begin{figure}[!ht]
	\centering
	\begin{tikzpicture}[scale=0.9, node distance=3cm, auto]
		\tikzstyle{ctxBlock} = [rectangle, draw=black, semithick, 
			text width=4.5em, text centered, rounded corners, minimum height=2em]
		\tikzstyle{machBlock} = [rectangle, draw=black, semithick, 
			text width=5.5em, text centered,  minimum height=2em]
		\tikzstyle{line} = [draw, thick,shorten >=1pt];
			\node [ctxBlock] (ctx0) {\small Ctx0\_Prm} ;
			\node [ctxBlock] (ctx1) [right = 1.9cm of ctx0] {\small Ctx1\_Prm} ;
			\node [machBlock] (abs) [below = 1cm of ctx0] {\small Operads\_Abs} ;
			\node [machBlock] (R1) [right = 1.5cm of abs] {\small Operads\_R1} ;
			\node [machBlock] (R2) [right = 1.5cm of R1] {\small Operads\_R2} ;
			\draw (ctx1)  edge[above , ->]  node {\tiny extends} (ctx0);
			\draw  (abs) edge[right , dashed, ->]  node {\tiny sees} (ctx0);
			\draw  (R1) edge[right , dashed,  ->]  node {\tiny sees} (ctx0);
			\draw  (R1)  edge[above, ->] node  {\tiny refines} (abs);
			\draw (R2)  edge[above, ->] node  {\tiny refines} (R1);
			\draw (R2) edge[right , dashed, ->]  node {\tiny sees} (ctx1);
	\end{tikzpicture}
	\captionof{figure}{Architecture of the Event-B development}
	\label{fig:genericArchi}	
\end{figure}
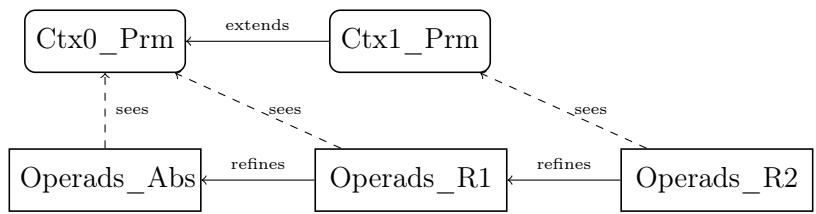

The architecture of the Event-B development is depicted in Fig.~\ref{fig:genericArchi}. 
First, we model a context (Ctx0\_Prm) that gathers the basic sets and parameters necessary to model the operads; 
this context is then extended with (Ctx1\_Prm) which introduce a set $X$  to particularize the inputs/outputs. 
The abstract machine (Operads\_Abs) contains the variables: 
\textit{myOperads} a set of operads, that will be eventually transformed or composed,
\textit{outOp} the outputs of each operad, 
\textit{arityOp} the arity function of each operad, and 
\textit{foliage} the growing foliage of an operad. These variables capture the structure of an operad, as it is seen from an abstract point of view, only the foliage is required; but we do not have yet the internal structuring of the foliage. This is done in the refinement. 
The events \texttt{newOperad} and \texttt{composeSeq} are defined in this first abstract machine (Operads\_Abs).

\paragraph{First refinement of the model (Operads\_R1)}
In the first refinement, we introduce the following variables: \textit{GHatOp}, a relation that gives the operad of any given argument at position \textit{ii};
\textit{hookOp} a partial function that gives if it exists, the operad in which another  one is hooked,   and \textit{gHookOp} a partial function that gives all the operads hooked in a given other one operad. The previous abstract events are refined accordingly.

\paragraph{Second refinement of the model (Operads\ R2)}
This refinement is used to introduce the set $X$, which specializes the input and output arguments of each operad.

We now introduce $X$ to decorate the inputs and outputs of the operads. The set $X$ can be seen as an abstraction of the cubes or disks (see Fig.~\ref{fig:cubeoperad}), where each input has its $x \in X$. It is just like adding details to the abstract model without changing its behaviour.
An Event-B refinement is used to model this addition of details into the previous  machine where the necessary structures was already introduced; they are then made more concrete, adding an $x$ to each input/output parameter of an operad. This is achieved, for instance regarding the input parameters of an operad (see \textsf{invr10} in Fig.~\ref{figure:M1_invariant}), with the partial injection (denoted with the operator $\pinj$) that associates an $x$ to each input argument:
$$ inOpX : myOperads \pfun (seqN(maxArgs) \pinj X)$$ 

For the consistency, the refinement technique requires to link abstract and concrete structures of the Event-B models. This is achieved for example with the following predicate (\ref{eq:absConc}) which expresses that all the operads that have input arguments in the abstract model  have arguments in the concrete model:
\begin{multline}
	\forall op. ((op \in myOperads \land op \in dom(inOp)) \implies 
	 (inOp(op) = dom(inOpX(op))) ) 
	\label{eq:absConc}
\end{multline}

\medskip

Through abstraction and refinements, we have implemented a complete development of an Event-B model designed as a correct-by-construction support for the mechanisation of operad usage.

	\section{Assessment and Exploiting the Proposed Development}
	We have yet conducted (through \texttt{Rodin/ProB}) several experiments with the current model.

\subsection{Assessement}
\label{section:assessment}

As stated in Sect. \ref{section:modelling}, the \textsf{Rodin} platform supports our development.  Model-checking (with ProB) is used to complement theorem proving available through the various provers of \textsf{Rodin}.  In combination with the \textsf{Rodin} provers, we used intensively the capabilities of ProB~\cite{DBLP:conf/fm/LeuschelB03} (model-checking, animation, disproving) to set up our model. Together with the model invariant, the specific properties SP1, SP2, SP3 where thoroughly analysed. 

\noindent
\begin{figure}[!ht]
	\centering
	\includegraphics*[width=0.6\textwidth]{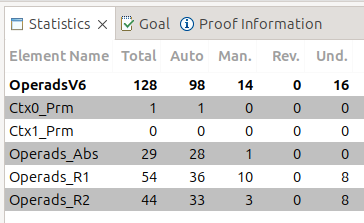}
	\caption{Reported result of proving with \textsf{Rodin}}
	\label{fig:resultTP1}	
\end{figure}

\paragraph{Proof statistics}
 Fig.~\ref{fig:resultTP1} depicts some proving statistics of the current Event-B project (the models as implemented with \textsf{Rodin}). The abstract model (named \texttt{Operads\_Abs}) is totally proved. It remains 8 proof obligations in the refined models (\texttt{Operads\_R1} and \texttt{Operads\_R2}). These PO are related either to the cardinality proofs, or to complicated typing (on intervals) in the formulation the properties SP$_i$.

\begin{figure}[!ht]
	\centering
\includegraphics*[width=0.8\textwidth]{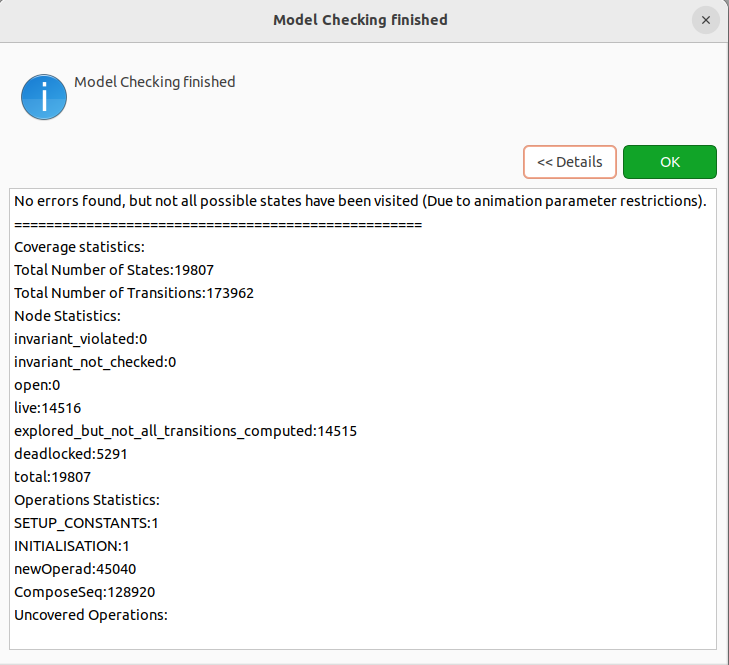}
\caption{Reported result of model checking with ProB}
\label{fig:resultMC1}	
\end{figure}

We have inspected the generated animation traces to analyse possible invariant violation, or to discover and correct errors in the model.
ProB was used to detect invariant violations and deadlock-freedom. Fig.~\ref{fig:resultMC1} depicts a statistics report of model-checking result. This shows that no invariant was violated after activating 45040 times the \textsf{newOperad} event and 128920 times the \textsf{composeSeq} event. The observed deadlocks are  situations in which no event can be fired due to simulation values (for instance, the cardinals of sets cardinals are limited for generating values); this leads to the machine parameters being reset and the analysis continuing.

\noindent

\subsection{Exploiting  the Proposed Development}
\label{section:howto}
The Event-B model we have built constitutes a basis for manipulating operad structures. The elementary operations of creation and sequential composition of operads are easily simulated with \textsf{Rodin} and \textsf{ProB}. 

\paragraph{Simulation} We have used \textsf{Rodin+ProB} to simulate the  proposed Event-B model with  arbitrary numbers of operads and arbitrary number of input arguments. This is done by  instantiating  the parameters in the Ctx0\_Prm context.

Similarly, we simulated arbitrary levels of composition of operads. The simulation was accompanied with model-checking (using \textsf{ProB}) of the model, in order to analyse the proposed mechanisation of operads.  

\paragraph{Extension} The current Event-B model can be extended with additional operations (various transformations) on operads by applying the Event-B refinement technique either on the Operads\_R1 machine or on the Operads\_R2 machine.  Specific properties can also be defined and added to the invariant to extend or to constrain  the model as needed.  

Mastering heterogeneity in complex systems can controlled through refinements. For example,  specific semantics or behaviour can be defined for composite operads and their arguments, as we did with the Operads\_R2 refinement. In general, the set $X$ can be interpreted or replaced by a semantic domain that denotes, for example, a vector space.

Similarly, the Event-B model can serve as a semantic foundation and reasoning system for modelling complex systems (natural or industrial)  at a higher level of abstraction.  



	\section{Conclusion}
	\label{section:conclusion}

We have considered mechanising operads in order to promote their use for modelling and analysing complex systems, using compositions of elementary objects.
While several theoretical works deal with operads, this work presents a practical implementation for operad algebras. It constitutes the basis of concrete library implementation, either for symbolic computation or for specific domain applications.

\paragraph{Summary of findings}
We have built a formal Event-B model as a basis for symbolic computations on operads. 
This model enables one to build n-ary operads and to compose them at any level by replacing an argument at a given position with another operad, resulting in a new operad.
The Event-B model provides a rigorous framework for the mechanisation of such algebraic structures. 
They can assist in the formal implementation of various applications such as the structuring or restructuring of software architectures, the description and operation of complex systems, the description or simulation of various evolving dynamic systems, etc.  
The expressive power, abstraction capabilities,  and the refinement techniques provided by Event-B allowed us to manipulate these operad structures in a practical manner. 
The structure of the proposed Event-B development may be adapted to extend the current work and address other challenging symbolic computation projects. 

\paragraph{Future work}
As part of our future work, we plan to construct Event-B theories of operads that will enable their widespread and practical use in various fields and applications.  First, the model currently proposed will be extended with commutativity and associativity relations \cite{AlgebraicOperadsLoday2012} defined as  operad axioms. 
This will facilitate, for example, the construction of and reasoning about systems modelled with operads.  Subsequently, various constructions and transformations of tree-structured systems can be endowed with, and supported by predefined theories.

	\medskip
	
	\medskip

	\noindent
	\paragraph{\textbf{Acknowledgement}} 
		\noindent 
		This work originated from fruitfull discussions with my colleague Johan L. Thanks Johan! My colleague Guillaume C. gave me a lot of encouragement. Thanks Guillaume! 

\bibliographystyle{plain}
\bibliography{./biblioOp-EB.bib}

\end{document}